\documentclass[conference]{IEEEtran}
\usepackage[T1]{fontenc}
\usepackage[latin9]{inputenc}
\usepackage{array}
\usepackage{bm}
\usepackage{multirow}
\usepackage{amsmath}
\usepackage{amsthm}
\usepackage{amssymb}
\usepackage{graphicx}
\usepackage[unicode=true,
 bookmarks=true,bookmarksnumbered=true,bookmarksopen=true,bookmarksopenlevel=1,
 breaklinks=false,pdfborder={0 0 0},pdfborderstyle={},backref=false,colorlinks=false]
 {hyperref}
\hypersetup{pdftitle={Your Title},
 pdfauthor={Your Name},
 pdfpagelayout=OneColumn, pdfnewwindow=true, pdfstartview=XYZ, plainpages=false}

\makeatletter

\providecommand{\tabularnewline}{\\}

\theoremstyle{plain}
\newtheorem{thm}{\protect\theoremname}

\usepackage[caption=false,font=footnotesize]{subfig}
\usepackage{diagbox}
\usepackage{algorithm}
\usepackage{algorithmic}
\usepackage{color}
\usepackage{bm}

\IEEEoverridecommandlockouts
\usepackage[left=0.625in, right=0.625in, bottom=1.00in, top=0.75in]{geometry}

\@ifundefined{showcaptionsetup}{}{%
 \PassOptionsToPackage{caption=false}{subfig}}
\usepackage{subfig}
\makeatother

\providecommand{\theoremname}{Theorem}

\begin{document}
\title{Blind Performance Prediction for Deep Learning Based Ultra-Massive MIMO Channel Estimation}
\author{\author{\IEEEauthorblockN{Wentao Yu$^\dagger$, Hengtao He$^\dagger$, Xianghao Yu$^\ddagger$, Shenghui Song$^\dagger$, \\ Jun Zhang$^\dagger$, \textit{Fellow, IEEE}, and Khaled B. Letaief$^\dagger$, {\textit{Fellow, IEEE}}} 	\IEEEauthorblockA{ 		$^\dagger$Dept. of Electronic and Computer Engineering, Hong Kong University of Science and Technology, Hong Kong\\	$^\ddagger$Dept. of Electrical Engineering, City University of Hong Kong, Hong Kong  \\	Email: $^\dagger$wyuaq@connect.ust.hk, $^\dagger$\{eehthe, eeshsong, eejzhang, eekhaled\}@ust.hk, $^\ddagger$alex.yu@cityu.edu.hk} \thanks{This work was supported by the Hong Kong Research Grants Council under Grant 16209622, 16212922 and 16212120, and by the Shenzhen Science and Technology Innovation Committee under Grant SGDX20210823103201006.}   } }
\maketitle
\begin{abstract}
Reliability is of paramount importance for the physical layer of wireless
systems due to its decisive impact on end-to-end performance. However,
the uncertainty of prevailing deep learning (DL)-based physical layer
algorithms is hard to quantify due to the black-box nature of neural
networks. This limitation is a major obstacle that hinders their practical
deployment. In this paper, we attempt to quantify the uncertainty
of an important category of DL-based channel estimators. An efficient
statistical method is proposed to make \textit{blind} predictions
for the mean squared error of the DL-estimated channel \textit{solely}
based on received pilots, without knowledge of the ground-truth channel,
the prior distribution of the channel, or the noise statistics. The
complexity of the blind performance prediction is low and scales only
linearly with the number of antennas. Simulation results for ultra-massive
multiple-input multiple-output (UM-MIMO) channel estimation with a
mixture of far-field and near-field paths are provided to verify the
accuracy and efficiency of the proposed method. 
\end{abstract}

\IEEEpeerreviewmaketitle{}

\section{Introduction\label{sec:Introduction}}

As the adoption of the 5-th generation (5G) wireless networks continues
to accelerate throughout the world, we are witnessing exciting global
research and development activities to formulate 6G as the next-generation
mobile communication network. In particular, 6G is envisioned to be
both a driver and a beneficiary of artificial intelligence (AI) \cite{2019Letaief}.
On one hand, ubiquitous high-speed wireless links in 6G lay the foundation
for emerging AI applications, such as edge intelligence, federated
learning, and extended reality \cite{2022Letaief}. On the other hand,
AI, in particular DL, facilitates the operation of increasingly complicated
wireless systems, and has already achieved great success in the physical
layer, such as data detection \cite{2020He}, hybrid precoding \cite{2022Ma,2020Zhang},
and channel estimation \cite{2022Yu,2017Borgerding,2022He}. 

However, the lack of reliability guarantee is a critical concern that
hinders the widespread deployment of DL-based physical layer design.
Traditional physical layer algorithms, such as the least squares,
have deeply rooted mathematical foundations and performance guarantees.
In contrast, DL-based methods are mostly treated as black boxes \cite{2022ShenTWC}.
Although they can learn from huge amounts of training data and achieve
the state-of-the-art performance, the over-parameterized and non-transparent
nature makes them hard to analyze. It is difficult to even have an
idea of whether DL-based methods are working well or not in practical
deployment. By taking channel estimation as an example, the commonly-adopted
mean squared error (MSE) metric requires knowledge of the ground-truth
channel, which is unavailable in deployment. Therefore, we can hardly
identify whether the DL-based channel estimators are making mistakes,
until the error has been propagated to higher levels and reveals itself
as, e.g., a decrease in the communication quality. This is certainly
undesirable for the physical layer which has stringent requirements
on reliability and robustness. 

Given such a dilemma, it is crucial to study how to quantify the uncertainty
of DL-based physical layer algorithms without the ground-truth data.
Such efforts not only serve as a safeguard so that traditional algorithms
can take over when mistakes are detected, but also provide valuable
insights for the design of the subsequent modules in the wireless
system. In addition, good performance criteria without the ground-truth
data can serve as unsupervised loss functions that enable online updates
of the DL-based physical layer algorithms \cite{2021Zheng}. In this
study, we focus on the channel estimation problem as a starting point. 

To realize these possibilities, we propose a \textit{blind} performance
prediction method for the (normalized) MSE of the estimated channel
\textit{without} any oracle information\footnote{By oracle, we refer to what cannot be known in practical deployment,
e.g., the ground-truth channel, the prior distribution, and the noise
statistics.}. The proposed method is compatible with an important category of
state-of-the-art DL-based channel estimators relying on approximate
message passing (AMP) and its variants \cite{2022Yu,2017Borgerding,2022He,2022Zou}.
The basic idea is to utilize the Gaussian error property of AMP-related
algorithms together with the Stein's unbiased risk estimator (SURE)
to predict the MSE performance \cite{1981Stein}. To enable the computation
of SURE, we also propose a principal component analysis (PCA)-based
method to estimate the noise level from a single instance of noisy
channel. Simulation results for channel estimation in UM-MIMO systems
are provided to verify the accuracy and efficiency of the proposed
method. To the best of the authors' knowledge, this is the first work
on the uncertainty quantification of DL-based channel estimation algorithms. 

Different from \cite{2021Gallyas-Sanhueza} which was proposed for
the thresholding-based sparse channel denoising problems, this paper
targets the more general compressed channel estimation problem, eliminates
the need of sparsity, and is tailored for DL-based methods. Unlike
the state evolution of AMP-related algorithms, which focuses on performance
analysis and requires the oracle information \cite{2022Zou}, this
paper instead targets blind performance prediction and does not require
any oracle information. 

\textit{Notation:} $\mathbf{A}^{T}$, $\mathbf{A}^{H}$, $\mathbf{A^{\dagger}}$,
and $\text{tr}(\mathbf{A})$ are the transpose, Hermitian, pseudo-inverse,
and trace of matrix $\mathbf{A}$, respectively. $\|\mathbf{a}\|_{p}$,
$(\mathbf{a})_{i}$, and $\max(\mathbf{a})$ are the $\ell_{p}$-norm,
the $i$-th element, and the maximum element of vector $\mathbf{a}$,
respectively. $|a|$ is the absolute value of scalar $a$. $\mathbf{B}=\text{blkdiag}(\mathbf{A}_{1},\mathbf{A}_{2},\ldots,\mathbf{A}_{n})$
returns a block diagonal matrix by aligning $\mathbf{A}_{1},\mathbf{A}_{2},\ldots,\mathbf{A}_{n}$
along the diagonal. $\mathbf{I}_{N}$ is an identity matrix of size
$N\times N$. $\mathcal{N}(\bm{\mu},\bm{\Sigma})$ and $\mathcal{CN}(\bm{\mu},\bm{\Sigma})$
are respectively the real-valued and complex-valued Gaussian distributions
with mean $\bm{\mu}$ and covariance $\bm{\Sigma}$. $\mathcal{U}(a_{1},a_{2})$
denotes a uniform distribution over the interval $[a_{1},a_{2}]$.
The divergence of a function $f:\mathbb{R}^{N}\rightarrow\mathbb{R}^{N}$
with respect to the input $\mathbf{x}$ is $\text{div}_{\mathbf{x}}\{f(\mathbf{x})\}\triangleq\frac{1}{N}\sum_{i=1}^{N}\frac{\partial(f(\mathbf{x}))_{i}}{\partial(\mathbf{x})_{i}}$. 

\section{System Model and Technical Background}

We consider the uplink channel estimation in millimeter-wave (mmWave)
or terahertz (THz) UM-MIMO systems as an example to present our proposal.
We first introduce the system and channel models. A category of DL-based
channel estimators relying on AMP-related algorithms is then discussed,
to which the proposed blind performance prediction method is applicable.
We later formulate the problem, and transform it to a more tractable
form via the Gaussian error property of AMP. 

\subsection{System and Channel Models}

Consider the uplink of a UM-MIMO system where the base station (BS)
is equipped with a uniform planar array (UPA) with $\sqrt{N}\times\sqrt{N}$
antennas and $N_{\text{RF}}$ radio frequency (RF) chains to serve
$K$ single-antenna user equipments (UEs). To enhance the energy efficiency,
a partially-connected hybrid analog-digital structure is adopted at
the BS, where each RF chain is connected with $N/N_{\text{RF}}$ antennas
\cite{2020Zhang}. The carrier frequency, carrier wavelength, antenna
spacing, and the speed of light are denoted as $f_{c}$, $\lambda_{c}$,
$d_{a}$, and $c$, respectively. 

The spatial-frequency channel $\tilde{\mathbf{h}}\in\mathbb{C}^{N\times1}$
of mmWave/THz UM-MIMO systems can be characterized by the superposition
of one line-of-sight (LoS) and $L-1$ non-LoS paths \cite{2022Yu},
i.e.,
\begin{equation}
\mathbf{\tilde{h}}=\sum_{l=1}^{L}\alpha_{l}\mathbf{a}\left(\phi_{l},\theta_{l},r_{l}\right)e^{-j2\pi f_{c}\tau_{l}},
\end{equation}
where $\alpha_{l}$, $\mathbf{a}\left(\phi_{l},\theta_{l},r_{l}\right)\in\mathbb{C}^{N\times1}$,
and $\tau_{l}$ denote the path loss, array response, and time delay
of the $l$-th path, respectively. The array response $\mathbf{a}\left(\phi_{l},\theta_{l},r_{l}\right)$
is determined by the azimuth angle-of-arrival (AoA) $\phi_{l}$, the
elevation AoA $\theta_{l}$, and the distance $r_{l}$ of the $l$-th
path. Due to the joint effects of large array aperture and small wavelength,
the near-field region becomes non-negligible in mmWave/THz UM-MIMO
systems \cite{2022Yu,2022Cao}. The array response, considering both
far-field (planar-wave) and near-field (spherical-wave) paths, is
given by
\begin{equation}
\mathbf{a}(\phi_{l},\theta_{l},r_{l})=\begin{cases}
\mathbf{a}^{\text{far}}(\phi_{l},\theta_{l},r_{l}), & \text{if }r_{l}>D_{\text{Rayleigh}},\\
\mathbf{a}^{\text{near}}(\phi_{l},\theta_{l},r_{l}) & \text{otherwise}.
\end{cases}
\end{equation}
Here $\mathbf{a}^{\text{far}}$ and $\mathbf{a}^{\text{near}}$ are
respectively the far- and near-field array responses, and $D_{\text{Rayleigh}}=\frac{2D^{2}}{\lambda_{c}}$
represents the Rayleigh distance, i.e., the boundary between the far
and near fields, with $D$ denoting the array aperture. The detailed
derivation of $\mathbf{a}^{\text{far}}$ and $\mathbf{a}^{\text{near}}$
can be found in \cite{2022Yu} and is omitted here for brevity. 

In the uplink training phase, the UEs send known pilots to the BS
for $Q$ time slots. Assuming that the widely-used orthogonal pilots
are adopted, we consider an arbitrary UE without loss of generality.
The received pilot signal $\mathbf{y}_{q}\in\mathbb{C}^{N_{\text{RF}}\times1}$
at the BS in time slot $q$ is given by
\begin{equation}
\mathbf{\tilde{y}}_{q}=\mathbf{W}_{\text{BB},q}^{H}\mathbf{W}_{\text{RF},q}^{H}\tilde{\mathbf{h}}s_{q}+\mathbf{W}_{\text{BB},q}^{H}\mathbf{W}_{\text{RF}}^{H}\tilde{\mathbf{n}}_{q},\label{eq:system-model}
\end{equation}
where $\mathbf{W}_{\text{BB},q}\in\mathbb{C}^{N_{\text{RF}}\times N_{\text{RF}}}$
is the digital combiner, $\mathbf{W}_{\text{RF},q}\in\mathbb{C}^{N\times N_{\text{RF}}}$
is the analog combiner, $s_{q}$ is the pilot symbol that is set as
1, and $\mathbf{\tilde{n}}_{q}\in\mathbb{C}^{N\times1}$ is the additive
white Gaussian noise (AWGN) vector following $\mathcal{CN}(\mathbf{0},\sigma_{\mathbf{\tilde{n}}}^{2}\mathbf{I}_{N})$.
Due to the partially-connected structure, we have that $\mathbf{W}_{\text{RF},q}=\text{blkdiag}(\mathbf{w}_{1,q},\mathbf{w}_{2,q},\ldots,\mathbf{w}_{N_{\text{RF}},q})$,
where each element of the vector $\mathbf{w}_{i,q}\in\mathbb{C}^{N/N_{\text{RF}}\times1}$
is picked from one-bit quantized angles, i.e., $(\mathbf{w}_{i,q})_{j}\in\sqrt{\frac{N_{\text{RF}}}{N}}\{\pm1\}$
\cite{2022Yu}. To whiten the effective noise, we set the digital
combiner as $\mathbf{W}_{\text{BB},q}=\mathbf{D}_{q}^{-1}$, where
$\mathbf{D}_{q}$ is an upper triangular matrix and $\mathbf{D}_{q}^{H}\mathbf{D}_{q}$
is the Cholesky decomposition of matrix $\mathbf{W}_{\text{RF},q}^{H}\mathbf{W}_{\text{RF},q}$.
The received pilot $\tilde{\mathbf{y}}=[\mathbf{\tilde{y}}_{1}^{T},\mathbf{\tilde{y}}_{2}^{T},\ldots,\mathbf{\tilde{y}}_{Q}^{T}]^{T}\in\mathbb{C}^{QN_{\text{RF}}\times1}$
of $Q$ time slots is $\mathbf{\tilde{y}}=\mathbf{\tilde{M}}\tilde{\mathbf{h}}+\mathbf{\tilde{n}}$,
where $\tilde{\mathbf{M}}=[(\mathbf{W}_{\text{BB},1}^{H}\mathbf{W}_{\text{RF},1}^{H})^{T},\ldots,(\mathbf{W}_{\text{BB},Q}^{H}\mathbf{W}_{\text{RF},Q}^{H})^{T}]^{T}\in\mathbb{C}^{QN_{\text{RF}}\times N}$,
and $\tilde{\mathbf{n}}\in\mathbb{C}^{QN_{\text{RF}}\times1}$ is
the whitened effective noise following $\mathcal{CN}(\mathbf{0},\sigma_{\mathbf{\tilde{n}}}^{2}\mathbf{I}_{QN_{\text{RF}}})$.
The system model can be transformed to its equivalent real-valued
form by stacking the real and imaginary parts \cite{2022Yu,2022He},
i.e.,
\begin{equation}
\mathbf{y}=\mathbf{M}\mathbf{h}+\mathbf{n},\label{eq:problem-formulation}
\end{equation}
with $\mathbf{y},\mathbf{n}\in\mathbb{R}^{2QN_{\text{RF}}\times1}$,
$\mathbf{M}\in\mathbb{R}^{2QN_{\text{RF}}\times2N}$, and $\mathbf{h}\in\mathbb{R}^{2N\times1}$.
The goal is to estimate the channel $\mathbf{h}$ from the received
signal $\mathbf{y}$ and the measurement matrix $\mathbf{M}$, where
the dimension of $\mathbf{y}$ is usually smaller than that of $\mathbf{h}$,
i.e., $QN_{\text{RF}}<N$, due to the requirement of low pilot overhead\footnote{In this paper, we focus on the compressed sensing-based channel estimation,
i.e., $QN_{\text{RF}}<N$. Nevertheless, both AMP-related algorithms
and the proposed blind performance prediction method can also be applied
when $QN_{\text{RF}}\geq N$.}. To solve the problem, AMP and related algorithms have attracted
great interest due to the low computational complexity and the tractable
error dynamics via the state evolution \cite{2022Zou,2017Ma}. Many
efficient DL-based estimators are developed based on them and achieved
promising performance \cite{2022Yu,2022He,2017Borgerding}, to which
the proposed blind prediction method can be effectively applied. 

\subsection{DL-Based Channel Estimators Relying on AMP\label{subsec:DL-Based-Channel-Estimators}}

AMP is an efficient algorithm for compressed sensing-based channel
estimation and works for an independent and identically distributed
(i.i.d.) sub-Gaussian measurement matrix $\mathbf{M}$. The per-iteration
update rules of AMP consist of a linear estimator (LE) and a non-linear
estimator (NLE) \cite{2022Zou}, given by
\begin{equation}
\begin{aligned}\begin{array}{cc}
 & \text{AMP-LE: }\end{array} & \begin{array}{cc}
\mathbf{r}^{(t)}=\mathbf{h}^{(t)}+\mathbf{M}^{T}(\mathbf{y}-\mathbf{M}\mathbf{h}^{(t)})+\mathbf{r}_{\text{Onsager}}^{(t)},\end{array}\\
\begin{array}{cc}
 & \text{AMP-NLE: }\end{array} & \begin{array}{cc}
\mathbf{h}^{(t+1)}=\eta_{t}(\mathbf{r}^{(t)}),\end{array}
\end{aligned}
\end{equation}
where the superscript $(t)$ indicates the $t$-th iteration, and
$\mathbf{r}_{\text{Onsager}}^{(t)}=\frac{N}{QN_{\text{RF}}}\text{div}_{\mathbf{r}^{(t)}}\{\eta_{t}(\mathbf{r}^{(t)})\}(\mathbf{r}^{(t-1)}-\mathbf{h}^{(t-1)})$
is the Onsager correlation term that ensures the input of the NLE
is the ground-truth channel $\mathbf{h}$ corrupted by AWGN, i.e.,
$\mathbf{r}^{(t)}=\mathbf{h}+\mathbf{e}^{(t)}$, where $\mathbf{e}^{(t)}\sim\mathcal{N}(\mathbf{0},\sigma_{\mathbf{e}^{(t)}}^{2}\mathbf{I}_{2N})$
holds, and $\eta_{t}(\cdot)$ is a Lipschitz-continuous function \cite{2022Zou}.
Orthogonal AMP (OAMP) is an extension that eliminates the Onsager
correlation and is compatible with a wider range of measurement matrices
\cite{2017Ma,2022Zou}. The per-iteration update rules of OAMP can
be summarized as 
\begin{equation}
\begin{aligned}\begin{array}{cc}
 & \text{OAMP-LE: }\end{array} & \begin{array}{cc}
\mathbf{r}^{(t)}=\mathbf{h}^{(t)}+\mathbf{W}^{(t)}(\mathbf{y}-\mathbf{M}\mathbf{h}^{(t)}),\end{array}\\
\begin{array}{cc}
 & \text{OAMP-NLE: }\end{array} & \begin{array}{cc}
\mathbf{h}^{(t+1)}=\eta_{t}(\mathbf{r}^{(t)}),\end{array}
\end{aligned}
\label{eq:OAMP}
\end{equation}
where the LE is constructed to be de-correlated and the NLE is designed
to be a divergence-free function \cite{2017Ma}. We say that the LE
is de-correlated when $\text{tr}(\mathbf{I}_{2N}-\mathbf{W}^{(t)}\mathbf{M})=0$
holds. The matrix $\mathbf{W}^{(t)}\in\mathbb{R}^{2N\times2QN_{\text{RF}}}$
is built upon either the matched filter, the pseudo-inverse, or the
linear minimum MSE (LMMSE) estimators \cite{2017Ma}. The divergence-free
NLE can be constructed from an arbitrary function by subtracting its
divergence \cite{2017Ma}. Similar to AMP, OAMP can also guarantee
that the input of the NLE equals $\mathbf{r}^{(t)}=\mathbf{h}+\mathbf{e}^{(t)}$
with $\mathbf{e}^{(t)}\sim\mathcal{N}(\mathbf{0},\sigma_{\mathbf{e}^{(t)}}^{2}\mathbf{I}_{2N})$,
i.e., the ground-truth channel $\mathbf{h}$ corrupted by AWGN \cite{2017Ma}. 

An important property of the AMP and OAMP is that the NLE can be interpreted
as a Gaussian noise denoiser \cite{2016Metzler}. Therefore, powerful
DL-based denoisers could be incorporated to enhance the performance.
Empirical results in the literature showed that the Gaussian error
property still holds even if black-box denoisers are utilized, such
as neural networks \cite{2017Borgerding,2016Metzler}. The proposed
blind performance predictor is established only based on the Gaussian
error property, which can be easily satisfied by a range of AMP-related
algorithms \cite{2022Zou} and their DL-based variants, indicating
its wide applicability. 
\begin{figure}[t]
\centering{}\includegraphics[width=0.35\textwidth]{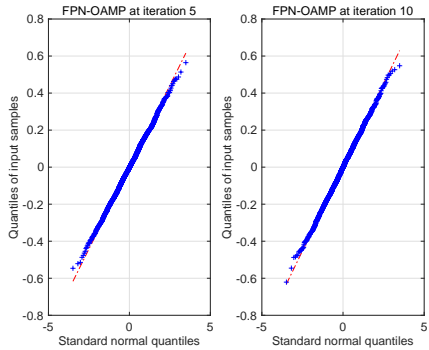}\caption{QQ plot of the NLE input error evaluated at iterations 5 and 10 of
the FPN-OAMP algorithm when SNR equals 15 dB. The plots show that
the error follows zero-mean Gaussian distribution due to the de-correlated
LE.\label{fig:QQ-plot-of} }
\end{figure}

As an example, we verify the Gaussian error property of our previously
proposed FPN-OAMP algorithm \cite{2022Yu,yu2022adaptive}. Specifically,
it adopts a de-correlated LE based on the pseudo-inverse matrix, i.e.,
$\mathbf{W}^{(t)}=\frac{2N}{\text{tr}(\mathbf{M}^{\dagger}\mathbf{M})}\mathbf{M}^{\dagger}$,
and a convolutional neural network-based NLE, i.e., $\eta_{t}(\cdot)=R_{\bm{\theta}}(\cdot)$,
whose weights $\bm{\theta}$ are identical in each iteration. The
update rules of FPN-OAMP are basically the same as (\ref{eq:OAMP})
with guaranteed convergence \cite{2022Yu}. The details of the dataset,
the training process, and the neural architecture are deferred to
Section \ref{sec:Simulation-Results}. In Fig. \ref{fig:QQ-plot-of},
we provide the quantile-quantile (QQ) plots of the NLE input error
of the FPN-OAMP algorithm evaluated at iterations 5 and 10 when the
received SNR is 15 dB. It can be observed that the quantiles fall
on the diagonal reference line, confirming that the NLE input error
is Gaussian distributed for DL-based channel estimators relying on
AMP-related algorithms. Similar results were also illustrated and
discussed in the literature \cite{2017Borgerding}. 

\subsection{Problem Formulation and Transformation}

The target of the blind performance prediction is the MSE of the estimated
channel $\mathbf{h}^{(t+1)}$ in each iteration, defined as
\begin{equation}
MSE^{(t)}=\|\mathbf{h}^{(t+1)}-\mathbf{h}\|_{2}^{2}=\|\eta_{t}(\mathbf{r}^{(t)})-\mathbf{h}\|_{2}^{2}.\label{eq:MSE}
\end{equation}
We cannot directly compute it since the ground-truth channel $\mathbf{h}$
is unknown in practical deployment. Instead, we want to derive a prediction
of it without using any oracle information. Based on the Gaussian
error property discussed before, we have
\begin{equation}
\mathbf{r}^{(t)}=\mathbf{h}+\mathbf{e}^{(t)},\label{eq:gaussian-noise}
\end{equation}
which is the ground-truth channel corrupted by some AWGN. Therefore,
the original problem can be transformed to predicting the MSE of a
denoising problem with the DL-based denoiser $\eta_{t}(\cdot)$, where
the ground-truth channel $\mathbf{h}$, the prior distribution of
the channel, and the variance of the noise $\sigma_{\mathbf{e}^{(t)}}^{2}$,
are all unknown. The only available information in (\ref{eq:gaussian-noise})
is the AWGN corrupted channel $\mathbf{r}^{(t)}$, which can be computed
solely based on the received signal $\mathbf{y}$ and the measurement
matrix $\mathbf{M}$. 

Based on (\ref{eq:gaussian-noise}), we notice that $\mathbf{r}^{(t)}$
can actually be interpreted as a Gaussian random variable, i.e., $\mathbf{r}^{(t)}\sim\mathcal{N}(\mathbf{h},\sigma_{\mathbf{e}^{(t)}}^{2}\mathbf{I}_{2N})$.
The function of the denoiser, $\eta_{t}(\cdot)$, is to estimate the
mean $\mathbf{h}$ of the Gaussian random variable $\mathbf{r}^{(t)}$.
The MSE of such an estimation can be predicted based on the seminal
work of Stein \cite{1981Stein}. In the following, we tackle the blind
performance prediction problem based on SURE \cite{1981Stein} and
PCA. 

\section{Blind Performance Prediction}

SURE is an unbiased estimate of the MSE of an estimator for the mean
of a multivariate Gaussian random variable. We first present the main
theorem of SURE using the notation from our setting, and then discuss
the difficulties in applying it for blind performance prediction and
the ways to resolve them. 
\begin{thm}[\cite{1981Stein,2007Blu}]
Given $\mathbf{r}^{(t)}$, $\mathbf{e}^{(t)}$ and $\eta_{t}(\cdot)$
as defined in (\ref{eq:gaussian-noise}), the $SURE^{(t)}$ corresponding
to $\eta_{t}(\mathbf{r}^{(t)})$ is a random variable specified by
the following equation, 
\begin{equation}
SURE^{(t)}\text{\!}=\!\|\eta_{t}(\mathbf{r}^{(t)})-\mathbf{r}^{(t)}\|_{2}^{2}-2N\sigma_{\mathbf{e}^{(t)}}^{2}\!+2\sigma_{\mathbf{e}^{(t)}}^{2}{\rm div}_{\mathbf{r}^{(t)}}(\eta_{t}(\mathbf{r}^{(t)})),\label{eq:SURE}
\end{equation}
which is an unbiased estimator of the MSE defined in (\ref{eq:MSE}),
i.e.,
\begin{equation}
\mathbb{E}_{\mathbf{e}^{(t)}}[SURE^{(t)}]=MSE^{(t)}.
\end{equation}
\end{thm}
\begin{IEEEproof}
Please refer to the proof of \cite[Theorem 1]{2007Blu}. 
\end{IEEEproof}
We refer to the three terms on the right-hand-side of (\ref{eq:SURE})
as the fidelity, variance, and divergence terms, respectively. The
theorem states that the expectation of $SURE^{(t)}$ is equal to $MSE^{(t)}$,
without the knowledge of the ground-truth channel $\mathbf{h}$. In
addition, thanks to the law of large numbers, if the dimension of
$\mathbf{r}^{(t)}$, i.e., $2N$, is sufficiently large, the variance
of $SURE^{(t)}$ will decrease in proportion to $1/2N$, and will
asymptotically converge to $MSE^{(t)}$ as the number of antennas
increases, i.e., $\lim_{N\rightarrow\infty}SURE^{(t)}=MSE^{(t)}$
\cite{2008Ramani}. For the considered UM-MIMO systems equipped with
thousands of antennas, using $SURE^{(t)}$ as a surrogate for $MSE^{(t)}$
can provide accurate blind performance prediction, as will be shown
in Section \ref{sec:Simulation-Results}. 

The difficulties in applying SURE for blind performance prediction
come from two aspects. First, the variance term $\sigma_{\mathbf{e}^{(t)}}^{2}$
is unknown, which should also be estimated from the AWGN corrupted
channel $\mathbf{r}^{(t)}$. Second, the divergence of the DL-based
NLE, $\eta_{t}(\cdot)$, can hardly be calculated in closed form since
it is not a component-wise function. We resolve the first difficulty
by using a PCA-based algorithm, while tackling the second issue by
an efficient Monte-Carlo method. Since the fidelity term can be directly
calculated based on $\mathbf{r}^{(t)}$ and $\eta_{t}(\cdot)$, we
mainly focus on computing the variance and divergence terms. 
\begin{figure*}[t]
\centering{}\subfloat[\label{fig:Array-of-subarray}]{
\centering{}\includegraphics[width=0.27\textwidth]{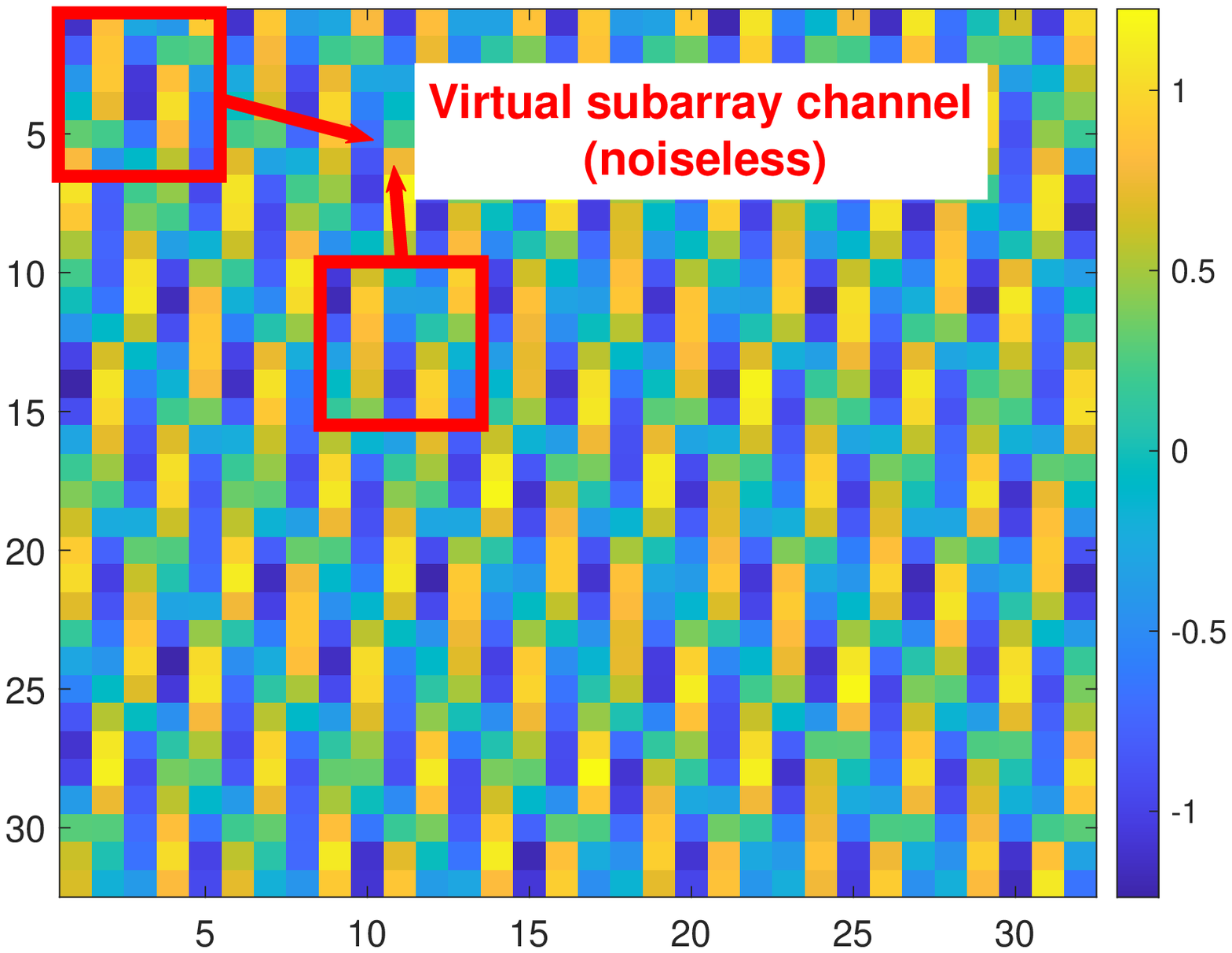}}\subfloat[\label{fig:Partially-connected}]{\centering{}\includegraphics[width=0.27\textwidth]{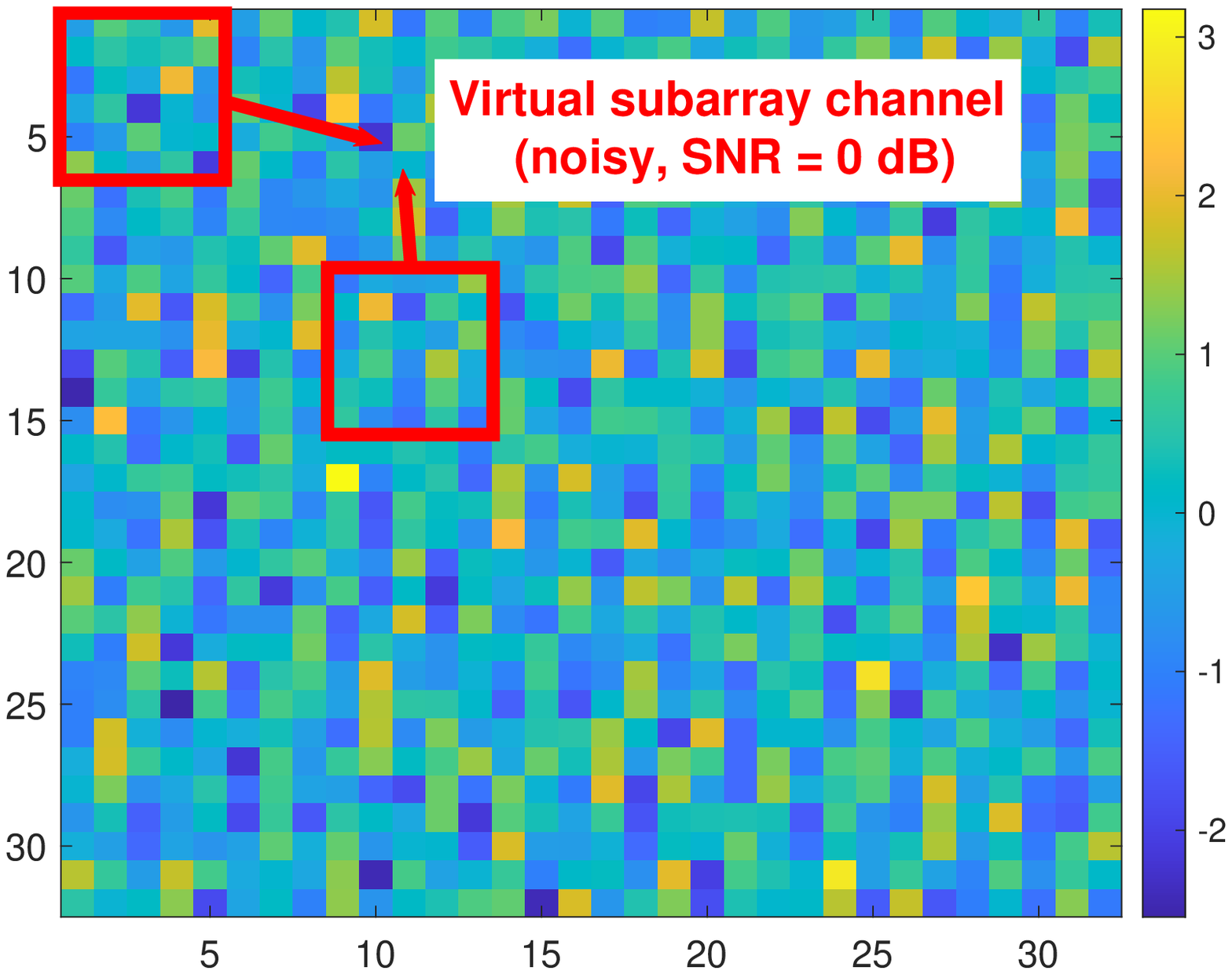}}\subfloat[\label{fig:Hybrid-field}]{\centering{}\includegraphics[width=0.27\textwidth]{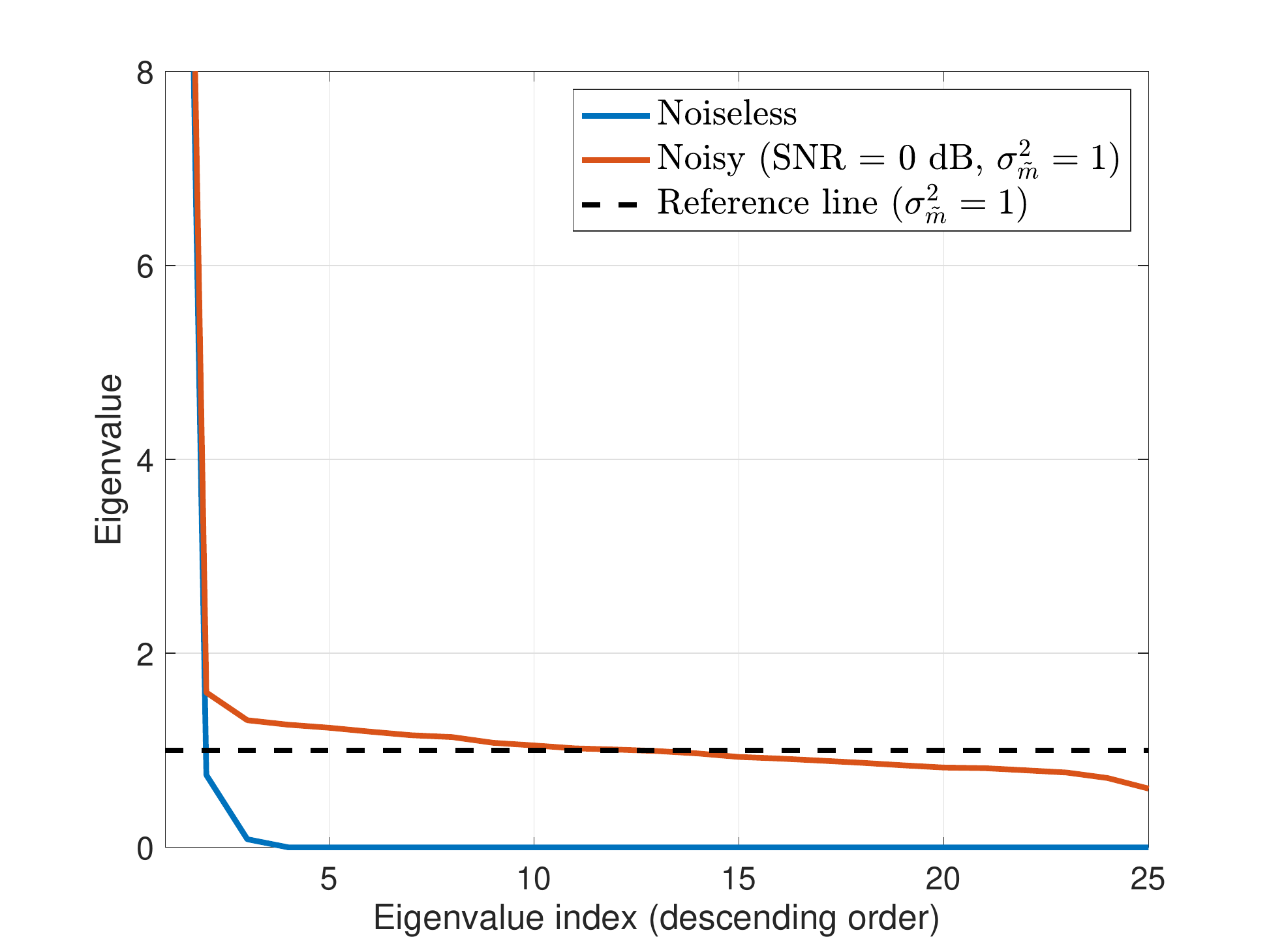}}\caption{(a) Heatmap of the real part of the noiseless channel. (b) Heatmap
of the real part of the noisy channel corrupted by AWGN with 0 dB
SNR ($\sigma_{\mathbf{\tilde{m}}}^{2}=1$). (c) Eigenvalues of the
covariance matrices of the noiseless and noisy virtual subarray channels
in descending order. The illustrated channel has $L=3$ paths. \label{fig:System-model}}
\end{figure*}

\subsection{The Variance Term\label{subsec:The-Variance-Term}}

The AMP and OAMP algorithms provide intuitive ways to track the variance
term $\sigma_{\mathbf{e}^{(t)}}^{2}$ \cite{2017Ma}. However, they
both depend on the statistics of the environmental noise, i.e., $\sigma_{\mathbf{\tilde{n}}}^{2}$,
which requires a dedicated training phase to estimate and may not
be accurately known in practice. In addition, the accuracy of these
methods is poor when DL-based components are involved, which will
be shown by simulations in Section \ref{sec:Simulation-Results}. 

We instead propose to estimate the noise level directly based on the
AWGN corrupted channel $\mathbf{r}^{(t)}\sim\mathcal{N}(\mathbf{h},\sigma_{\mathbf{e}^{(t)}}^{2}\mathbf{I}_{2N})$,
which is the only available information in (\ref{eq:gaussian-noise}).
Since the mean, i.e., the ground-truth channel $\mathbf{h}$, is \textit{unknown},
the problem is essentially to estimate the noise variance $\sigma_{\mathbf{e}^{(t)}}^{2}$
from a \textit{single} realization of a multivariate Gaussian distribution
with \textit{unknown} mean. This is \textit{intractable} if the channel
$\mathbf{h}$ does not have any structure. 

To exploit the structural information, previous work proposed to transform
the AWGN corrupted channel $\mathbf{r}^{(t)}$ into its sparse angular-domain
representation and estimate the noise variance with median absolute
deviation \cite{2021Gallyas-Sanhueza}. However, for a practical UM-MIMO
channel with both far- and near-field paths, the angular-domain channel
is no longer sparse \cite{2022Yu}, which greatly degrades the accuracy
of \cite{2021Gallyas-Sanhueza}. Therefore, we need to exploit other
structural information that generally holds for the UM-MIMO channel.
Specifically, we turn to the low-dimensional structure of the channel
due to the limited number of paths and adopt a PCA-based method as
shown in \textbf{Algorithm 1}. 

To explain the ideas behind, we generate a noisy UM-MIMO channel with
$L=3$ paths and $N=1024$ antennas with $\mathbf{\tilde{h}}_{\text{noisy}}=\mathbf{\tilde{h}}+\mathbf{\tilde{m}}$
and $\mathbf{\tilde{m}}\sim\mathcal{CN}(\mathbf{0},\sigma_{\mathbf{\tilde{m}}}^{2}\mathbf{I}_{N})$.
We set $\sigma_{\mathbf{\tilde{m}}}^{2}=1$ such that the SNR, defined
as $1/\sigma_{\mathbf{\tilde{m}}}^{2}$, equals 0 dB. We reshape both
$\mathbf{\tilde{h}}_{\text{noisy}}$ and $\mathbf{\tilde{h}}$ into
$\sqrt{N}\times\sqrt{N}$ matrices and respectively plot their real
parts, as shown in Fig. \ref{fig:System-model}(a)(b). We then explain
the algorithm line by line with this visualized example. 

In lines 3 and 4, we take multiple virtual subarray channels (VSCs)
from the UM-MIMO channel using a sliding window, as shown by red squares
in Fig. \ref{fig:System-model}(a)(b). Since the noise is white Gaussian,
estimating the noise level for VSCs is equivalent to estimating that
for the original channel. Therefore, we resort to the statistical
properties of the VSCs instead. 

In lines 5 to 7, we first calculate the covariance matrix $\bm{\Sigma}$
of the noisy VSCs, and then compute the eigenvalues of $\bm{\Sigma}$
and list them in descending order. In this process, each VSC is treated
as a data point in the PCA, and the eigenvalues of $\bm{\Sigma}$
reflect the strength of the different components. We plot the eigenvalues
of the clean and noisy channels in Fig. \ref{fig:System-model}(c).
One can see that the clean channel only has $L=3$ non-zero eigenvalues,
equal to the number of paths. By contrast, the eigenvalues of the
noisy channel are long-tailed and surround the variance of the additive
noise, as shown by the reference line. 

In lines 8 to 14, we split the eigenvalues into the principal and
redundant dimensions, and estimate the noise level based on the latter,
because it was proved in \cite[Lemma 1]{2015Chen} that the redundant
eigenvalues follow $\mathcal{N}(\sigma_{\mathbf{\tilde{m}}}^{2}/2,\sigma_{\mathbf{\tilde{m}}}^{4}/2s)$
when $s$ is large enough. However, the problem is that the range
of the principal dimensions, i.e., the number of paths $L$, is unknown
in practical deployment. We adopt an iteration to split the dimensions,
and stop the iteration when the mean and median of the redundant eigenvalues
in the current iteration are found equal \cite{2015Chen}. Then the
median is used to estimate the noise level. 

In blind performance prediction, $\mathbf{r}^{(t)}$ is the same as
the noisy channel mentioned above. The algorithm can thus be similarly
applied to calculate the variance term $\sigma_{\mathbf{e}^{(t)}}^{2}$. 

\begin{algorithm}[!tbp]
\caption{PCA-based noise level estimation}
\begin{algorithmic}[1]
\STATE {\bf Input:} AWGN corrupted channel $\tilde{\mathbf{h}}_{\text{noisy}} \in \mathbb{C}^{N \times 1}$, size of the virtual subarray $\sqrt{d} \times \sqrt{d}$\\
\STATE {\bf Output:} Estimated noise level $\hat{\sigma}_{\tilde{\mathbf{m}}}$ \\
\STATE \textcolor{black}{Reshape $\tilde{\mathbf{h}}_{\text{noisy}} \in \mathbb{C}^{N \times 1}$ into the equivalent real-valued tensor form $\mathbf{H}_{\text{noisy}} \in \mathbb{R}^{\sqrt{N} \times \sqrt{N} \times 2}$} \\
\STATE \textcolor{black}{Decompose $\mathbf{H}_{\text{noisy}}$ into virtual subarray channels $\{\mathbf{h}_{\text{noisy},t} \in \mathbb{R}^{2d \times 1}\}_{t=1}^{s}$ by sliding window, with $s=(\sqrt{N}-\sqrt{d}+1)^2$} \\
\STATE \textcolor{black}{Calculate the mean vector of the virtual subarray channels, i.e., $\bm{\mu}=\frac{1}{s} \sum_{t=1}^{s}\mathbf{h}_{\text{noisy},t} \in \mathbb{R}^{2d \times 1}$} \\
\STATE \textcolor{black}{Calculate the covariance of the virtual subarray channels, i.e., $\bm{\Sigma}=\frac{1}{s}\sum_{t=1}^{s}(\mathbf{h}_{\text{noisy},t}-\bm{\mu})(\mathbf{h}_{\text{noisy},t}-\bm{\mu})^T \in \mathbb{R}^{2d \times 2d}$} \\
\STATE \textcolor{black}{Calculate the eigenvalues $\{\lambda_i\}_{i=1}^{2d}$ of the covariance $\bm{\Sigma}$, and sort them in descending order, i.e., $\lambda_1 \geq \lambda_2 \geq \ldots \geq \lambda_{2d}$} \\
\STATE \textcolor{black}{Iteratively split the eigenvalues $\{\lambda_i\}_{i=1}^{2d}$ into the principal and the redundant dimensions, and estimate the noise level based on the eigenvalues of the redundant dimensions: } \\
\STATE {\bf for} $i = 1:2d$ {\bf do}
\STATE\hspace{\algorithmicindent} $\tau = \frac{1}{2d-i+1}\sum_{j=i}^{2d} \lambda_j$
\STATE\hspace{\algorithmicindent} {\bf if} $\tau$ is the median of $\{\lambda_j\}_{j=i}^{2d}$ {\bf then}
\STATE\hspace{\algorithmicindent}\hspace{\algorithmicindent}\hspace{\algorithmicindent} $\hat{\sigma}_{\tilde{\mathbf{m}}}=\sqrt{2\tau}$ and {\bf break} 
\STATE\hspace{\algorithmicindent} {\bf end if }
\STATE {\bf end for}
\end{algorithmic}
\end{algorithm}

\subsection{The Divergence Term}

As defined in Section \ref{sec:Introduction}, the divergence needs
to take into account the effect of each element of the input vector.
However, the DL-based NLE, i.e., $\eta_{t}(\cdot)$, is not a component-wise
function, which makes closed-form computation of the divergence term
infeasible. In \cite{2008Ramani}, the authors proved that for bounded
function $\eta_{t}(\cdot)$, the divergence can be expressed as
\begin{equation}
\text{div}_{\mathbf{r}^{(t)}}(\eta_{t}(\mathbf{r}^{(t)}))=\lim_{\epsilon\to0}\mathbb{E}_{\bm{\delta}}[\frac{\bm{\delta}^{T}}{\epsilon}(\eta_{t}(\mathbf{r}^{(t)}+\epsilon\bm{\delta})-\eta_{t}(\mathbf{r}^{(t)}))],\label{eq:MC-SURE}
\end{equation}
where $\bm{\delta}\sim\mathcal{N}(\mathbf{0},\mathbf{I}_{2N})$ is
a Gaussian distributed random variable independent of $\mathbf{r}^{(t)}$,
and $\epsilon\bm{\delta}$ is a slight perturbation to the input of
the function $\eta_{t}(\cdot)$. This is a powerful result compatible
with the considered DL-based NLE, since the expression in (\ref{eq:MC-SURE})
does not require the explicit expression of $\eta_{t}(\cdot)$, but
rather the output of it. The remaining difficulty is that expectation
also does not have a closed form and cannot be computed directly.
Thanks to the law of large numbers, it can be efficiently approximated
via Monte-Carlo sampling. Owing to the high dimensionality of the
UM-MIMO channel, a single sample of $\bm{\delta}$ is good enough
to approximate the expectation in (\ref{eq:MC-SURE}) by setting $\epsilon$
as a small value \cite{2008Ramani}. We use $\epsilon=\frac{\max(\mathbf{r}^{(t)})}{100}$
throughout this work. Overall, the approximation can be computed by
\begin{equation}
\text{div}_{\mathbf{r}^{(t)}}(\eta_{t}(\mathbf{r}^{(t)}))\thickapprox\frac{\bm{\delta}^{T}}{\epsilon}(\eta_{t}(\mathbf{r}^{(t)}+\epsilon\bm{\delta})-\eta_{t}(\mathbf{r}^{(t)})),
\end{equation}
which is easy to implement and provides competitive empirical performance
according to the simulations in Section \ref{sec:Simulation-Results}. 

\subsection{Complexity Analysis}

We separately analyze the complexity of the fidelity, variance, and
divergence terms. For the fidelity term, the complexity of computing
the $\ell_{2}$-norm is $\mathcal{O}(N)$. For the variance term,
we analyze the complexity of \textbf{Algorithm 1} step by step. The
complexity of the virtual subarray decomposition in line 4 is $\mathcal{O}(sd)$.
Computing the mean and covariance in lines 5 and 6 costs $\mathcal{O}(sd)$
and $\mathcal{O}(sd^{2})$, respectively. The complexity of the eigenvalue
decomposition in line 7 is $\mathcal{O}(d^{3})$, while sorting the
eigenvalues costs $\mathcal{O}(d^{2})$. The eigenvalue splitting
and the noise level estimation from lines 8 to 14 costs $\mathcal{O}(d^{2})$.
Since in UM-MIMO systems $s\thickapprox N$ holds, the complexity
of calculating the variance term is roughly $\mathcal{O}(Nd^{2}+d^{3})$.
For the divergence term, the complexity lies in an additional forward
propagation of the DL-based NLE $\eta_{t}(\cdot)$. This incurs a
constant complexity once $\eta_{t}(\cdot)$ is determined, which we
denote as $\mathcal{O}(p)$ \cite{2022Yu}. Therefore, the total complexity
of the blind performance prediction method is $\mathcal{O}(Nd^{2}+d^{3}+p)$,
which scales only linearly with the number of antennas. This makes
the algorithm friendly with UM-MIMO systems with potentially thousands
of antennas. 

\section{Simulation Results\label{sec:Simulation-Results}}

\subsection{Simulation Setup}

In the following, we verify the accuracy of the proposed blind performance
prediction method with the uplink channel estimation in mmWave UM-MIMO\footnote{In the literature, UM-MIMO usually refers to systems equipped with
1024 or more antennas. Nevertheless, the proposed blind performance
prediction is also observed to work well with moderate-sized massive
MIMO systems with, e.g., 128 or 256 antennas. Theoretical analysis
on the impact of the array size to the prediction accuracy is left
as future work. } systems based on the FPN-OAMP algorithm \cite{2022Yu}. We first
introduce the system settings and the details of the algorithm. The
main system parameters are set as $N=1024$, $N_{\text{RF}}=4$, $L=3$,
$f_{c}=14$ GHz, $d_{a}=0.5\lambda_{c}$, $Q=128$, $\theta_{l}\sim\mathcal{U}(-\pi/2,\pi/2)$,
and $\phi_{l}\sim\mathcal{U}(-\pi,\pi)$. In this case, the Rayleigh
distance $D_{\text{Rayleigh}}$ is about 20 meters. We set the path
length as $r_{l}\sim\mathcal{U}(5,30)$ meters to model the mixture
of far-field and near-field paths contained in the UM-MIMO channel.
In addition, for the LoS path, we set the path loss as $\alpha_{l}\sim\mathcal{CN}(0,1)$,
while for the non-LoS paths, the path loss is set as $\alpha_{l}\sim\mathcal{CN}(0,0.1)$.
We normalize the channel $\mathbf{\tilde{h}}$ such that $\|\mathbf{\tilde{h}}\|_{2}^{2}=N$,
and define the received SNR as $1/\sigma_{\mathbf{\tilde{n}}}^{2}$. 

Following Section \ref{subsec:DL-Based-Channel-Estimators}, the DL-based
NLE $\eta_{t}(\cdot)=R_{\bm{\theta}}(\cdot)$ of FPN-OAMP \cite{2022Yu}
first reshapes $\mathbf{r}^{(t)}$ into tensor form $\mathbf{R}^{(t)}\in\mathbb{R}^{\sqrt{N}\times\sqrt{N}\times2}$,
and then passes it to a $3\times3$ convolution (Conv) layer to upsample
the input to 64 feature maps. These feature maps further go through
a ResNet-like structure consisting of three residual blocks (RBs)
\cite{2016He}. Each RB constitutes two $3\times3$ Conv layers with
64 feature maps, which are followed by ReLU activation and layer normalization
\cite{2016Ba}. The RBs are further followed by two $1\times1$ Conv
layers with 2 feature maps, whose output is reshaped back to vector
form as $\mathbf{h}^{(t+1)}$. 

We generate the training, validation, and testing datasets with 20000,
5000, and 5000 samples, respectively. The presented results are tested
based on the testing dataset. FPN-OAMP is trained for 150 epochs based
on the normalized MSE (NMSE) loss function using Adam optimizer. The
minibatch size is 128 and the initial learning rate is 0.001. The
learning rate is decayed by half at the end of every 40 epochs. During
the training, validation, and testing processes, we set the maximum
number of iterations of FPN-OAMP as 15. Other more detailed training
settings that are omitted here can be found in \cite{2022Yu}. 

\subsection{Noise Level Estimation}

The accuracy of the noise level estimation can affect both the variance
and the divergence terms in SURE, and is thus critical for the blind
performance prediction. We provide the simulation results to verify
the performance of the proposed PCA-based method. We first generate
the complex-valued AWGN corrupted channels by using $\mathbf{\tilde{h}}_{\text{noisy}}=\mathbf{\tilde{h}}+\mathbf{\tilde{m}}$
with $\mathbf{\tilde{m}}\sim\mathcal{CN}(\mathbf{0},\sigma_{\mathbf{\tilde{m}}}^{2}\mathbf{I}_{N})$,
and then transform it into the equivalent real-valued form, i.e.,
$\mathbf{\mathbf{h}}_{\text{noisy}}=\mathbf{\mathbf{h}}+\mathbf{\mathbf{m}}$.
We estimate the noise level $\sigma_{\mathbf{\tilde{m}}}$ from the
noisy channel $\mathbf{\tilde{h}}_{\text{noisy}}$ or $\mathbf{h}_{\text{noisy}}$.
The estimation is denoted by $\hat{\sigma}_{\mathbf{\tilde{m}}}$.
The SNR is defined as $1/\sigma_{\mathbf{\tilde{m}}}^{2}$. Two benchmarks
are compared: 
\begin{table}[t]
\begin{centering}
\caption{Performance of the noise level estimation \label{tab:Performance-of-the}}
\par\end{centering}
\centering{}%
\begin{tabular}{c|c|>{\centering}p{0.9cm}|>{\centering}p{0.7cm}|>{\centering}p{0.7cm}|c}
\hline 
$\sigma_{\mathbf{\tilde{m}}}$ (SNR) & Method & Bias & Std & RMSE & Runtime (s)\tabularnewline
\hline 
\multirow{3}{*}{0.5623 (5 dB)} & Oracle & 0.0001 & 0.0089 & 0.0089 & 0.0001\tabularnewline
\cline{2-6} \cline{3-6} \cline{4-6} \cline{5-6} \cline{6-6} 
 & \textbf{Proposed} & 0.0038 & 0.0128 & 0.0134 & 0.0039\tabularnewline
\cline{2-6} \cline{3-6} \cline{4-6} \cline{5-6} \cline{6-6} 
 & Sparsity & 0.0396 & 0.0233 & 0.0459 & 0.0007\tabularnewline
\hline 
\multirow{3}{*}{0.1778 (15 dB)} & Oracle & 0.0001 & 0.0027 & 0.0027 & 0.0001\tabularnewline
\cline{2-6} \cline{3-6} \cline{4-6} \cline{5-6} \cline{6-6} 
 & \textbf{Proposed} & 0.0015 & 0.0039 & 0.0042 & 0.0039\tabularnewline
\cline{2-6} \cline{3-6} \cline{4-6} \cline{5-6} \cline{6-6} 
 & Sparsity & 0.0367 & 0.0207 & 0.0422 & 0.0007\tabularnewline
\hline 
\multirow{3}{*}{0.0562 (25 dB)} & Oracle & <0.0001 & 0.0009 & 0.0009 & 0.0001\tabularnewline
\cline{2-6} \cline{3-6} \cline{4-6} \cline{5-6} \cline{6-6} 
 & \textbf{Proposed} & 0.0004 & 0.0012 & 0.0013 & 0.0038\tabularnewline
\cline{2-6} \cline{3-6} \cline{4-6} \cline{5-6} \cline{6-6} 
 & Sparsity & 0.0391 & 0.0237 & 0.0457 & 0.0007\tabularnewline
\hline 
\end{tabular}
\end{table}

\begin{itemize}
\item \textit{Sparsity-based method}\footnote{Since this benchmark was originally proposed for the complex Gaussian
distribution \cite{2021Gallyas-Sanhueza}, we test it using the complex-valued
noisy channel $\mathbf{\tilde{h}}_{\text{noisy}}$. }: As said in Section \ref{subsec:The-Variance-Term} \cite{2021Gallyas-Sanhueza}. 
\item \textit{Oracle bound}\footnote{This is the minimum variance unbiased estimator of the oracle setting. }:
Assume that the ground-truth channel $\mathbf{h}$ is known, and estimate
$\sigma_{\mathbf{\tilde{m}}}$ directly from $\mathbf{m}=\mathbf{\mathbf{h}}_{\text{noisy}}-\mathbf{h}$
by the sample standard deviation, i.e., $\hat{\sigma}_{\mathbf{\tilde{m}}}=\sqrt{\frac{1}{N}{\textstyle {\scriptstyle \sum_{i=1}^{2N}}}(\mathbf{m})_{i}^{2}}$. 
\end{itemize}
In Table \ref{tab:Performance-of-the}, we present the performance
of the noise level estimation with different noise levels. The bias,
the standard deviation (std), and the root MSE (RMSE) of the noise
level estimation are defined as $\mathbb{E}[|\sigma_{\mathbf{\tilde{m}}}-\mathbb{E}[\hat{\sigma}_{\mathbf{\tilde{m}}}]|]$,
$\sqrt{\mathbb{E}[(\hat{\sigma}_{\mathbf{\tilde{m}}}-\mathbb{E}[\hat{\sigma}_{\mathbf{\tilde{m}}}])^{2}]}$,
and $\sqrt{\mathbb{E}[(\sigma_{\mathbf{\tilde{m}}}-\hat{\sigma}_{\mathbf{\tilde{m}}})^{2}]}$,
respectively. The bias and std respectively reflect the accuracy and
the robustness of an estimator, while the RMSE reflects the overall
performance. The virtual subarray size of the proposed method is set
as $5\times5$, i.e., $d=25$. The runtime is tested using Matlab
on Intel Core i7-9750H CPU. 

From Table \ref{tab:Performance-of-the}, we observe that the proposed
PCA-based method is highly accurate and robust. The RMSE performance
is close to the oracle bound even though it does not have knowledge
of the ground-truth channel. However, the performance of the sparsity-based
method \cite{2021Gallyas-Sanhueza} is much worse, since the hybrid-field
UM-MIMO channel does not satisfy the angular-domain sparsity \cite{2022Yu}.
On the contrary, the proposed PCA-based method does \textit{not} rely
on channel sparsity, and is thus applicable to a much broader class
of channel models of practical interest. In addition, the runtime
of the proposed method, though higher than the sparsity-based one,
is within only a few milliseconds. 

\subsection{Blind Performance Prediction}

Since in practice we are more interested in normalized MSE (NMSE),
it is presented instead by normalizing the predicted MSE with $\|\mathbf{\mathbf{h}}\|_{2}^{2}$.
If $\|\mathbf{\mathbf{h}}\|_{2}^{2}$ is unknown, we can approximate
it by $\|\mathbf{r}^{(t)}\|_{2}^{2}-2N\sigma_{\mathbf{e}^{(t)}}^{2}$,
as suggested in \cite{2021Gallyas-Sanhueza}. This can provide almost
the same predicted NMSE as using the true channel power. We adopt
this approximation in the following results. 
\begin{figure}[t]
\centering{}\includegraphics[width=0.35\textwidth]{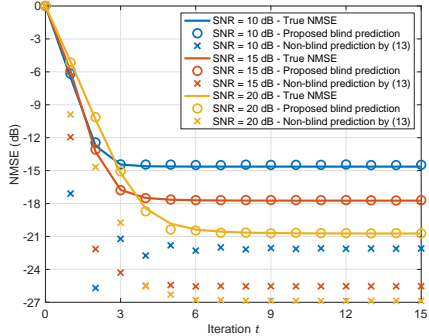}\caption{Iterative behavior of the FPN-OAMP algorithm in terms of the true
NMSE and predicted NMSE by different methods. \label{fig:Iterative-behavior-of}}
\end{figure}

To the best of our knowledge, this is the first attempt in this direction,
and we can hardly find relevant baselines to compare with. For the
sake of completeness, we provide a comparison with a \textit{non-blind}
and intuitive MSE performance prediction method commonly adopted in
previous work \cite{2020He,2017Ma}, i.e.,
\begin{equation}
\widehat{MSE}^{(t)}=2N\max\{\frac{\|\mathbf{y}-\mathbf{M}\mathbf{h}^{(t)}\|_{2}^{2}-QN_{\text{RF}}\sigma_{\mathbf{\tilde{n}}}^{2}}{\text{tr}(\mathbf{M}^{T}\mathbf{M})},\xi\},\label{eq:non-blind-prediction}
\end{equation}
which requires the noise statistics, i.e., $\sigma_{\mathbf{\tilde{n}}}^{2}$
(oracle information). The predicted NMSE is obtained similarly by
the normalization. This predictor is dependent on two much stronger
assumptions on the distribution of $\mathbf{\mathbf{r}}^{(t)}-\mathbf{h}$
and $\mathbf{h}^{(t)}-\mathbf{h}$ \cite{2020He,2017Ma}, whose detailed
explanation can be found in \cite[Appendix B]{2020He}. However, the
DL-based NLE can often break these assumptions and make the predictor
over-optimistic, or even return negative values for the predicted
MSE \cite{2017Ma}. To avoid unreasonable prediction, we set a floor
value of $\xi=0.001$ according to \cite{2017Ma}. 

In Fig. \ref{fig:Iterative-behavior-of}, we plot the true NMSE and
the predicted NMSE as a function of the number of iterations $t$.
The true NMSE is plotted with full line. The predicted NMSE by our
blind method and the intuitive method using (\ref{eq:non-blind-prediction})
is respectively plotted by circle and cross. It is observed that the
predicted NMSE by our method agrees very well with the true NMSE,
even though it is computed without using any oracle information. This
confirms the accuracy of the proposed blind performance prediction
method. By contrast, the prediction from the intuitive method using
(\ref{eq:non-blind-prediction}) is far away from the true performance.
Furthermore, the runtime of the proposed method is low, taking only
0.031 seconds when tested in Pytorch on Intel Core i7-9750H CPU. 

\section{Conclusion and Future Work}

This work proposed an efficient blind performance prediction method
for DL-based UM-MIMO channel estimation. Accurate prediction can be
achieved solely based on the received pilots, with no need of the
ground-truth channel or any other oracle information. The predicted
performance is useful for quantifying the uncertainty and identifying
the potential risk of DL-based algorithms, which can serve as valuable
design guidelines in practical deployment. For the future work, it
is interesting to investigate how DL-based methods can adapt themselves
when poor performance is detected based on the proposed framework. 

\bibliographystyle{IEEEtran}
\bibliography{references}

\begin{thebibliography}{10}
\providecommand{\url}[1]{#1}
\csname url@samestyle\endcsname
\providecommand{\newblock}{\relax}
\providecommand{\bibinfo}[2]{#2}
\providecommand{\BIBentrySTDinterwordspacing}{\spaceskip=0pt\relax}
\providecommand{\BIBentryALTinterwordstretchfactor}{4}
\providecommand{\BIBentryALTinterwordspacing}{\spaceskip=\fontdimen2\font plus
\BIBentryALTinterwordstretchfactor\fontdimen3\font minus
  \fontdimen4\font\relax}
\providecommand{\BIBforeignlanguage}[2]{{%
\expandafter\ifx\csname l@#1\endcsname\relax
\typeout{** WARNING: IEEEtran.bst: No hyphenation pattern has been}%
\typeout{** loaded for the language `#1'. Using the pattern for}%
\typeout{** the default language instead.}%
\else
\language=\csname l@#1\endcsname
\fi
#2}}
\providecommand{\BIBdecl}{\relax}
\BIBdecl

\bibitem{2019Letaief}
K.~B. Letaief, W.~Chen, Y.~Shi, J.~Zhang, and Y.-J.~A. Zhang, ``The roadmap to
  {6G}: {AI} empowered wireless networks,'' \emph{IEEE Comm. Mag.}, vol.~57,
  no.~8, pp. 84--90, Aug. 2019.

\bibitem{2022Letaief}
K.~B. Letaief, Y.~Shi, J.~Lu, and J.~Lu, ``Edge artificial intelligence for
  {6G}: Vision, enabling technologies, and applications,'' \emph{IEEE J. Sel.
  Areas Commun.}, vol.~40, no.~1, pp. 5--36, Jan. 2022.

\bibitem{2020He}
H.~He, C.-K. Wen, S.~Jin, and G.~Y. Li, ``Model-driven deep learning for {MIMO}
  detection,'' \emph{IEEE Trans. Signal Process.}, vol.~68, pp. 1702--1715,
  Feb. 2020.

\bibitem{2022Ma}
Y.~Ma, Y.~Shen, X.~Yu, J.~Zhang, S.~H. Song, and K.~B. Letaief, ``Learn to
  communicate with neural calibration: Scalability and generalization,''
  \emph{IEEE Trans. Wireless Commun.}, vol.~21, no.~11, pp. 9947--9961, Nov.
  2022.

\bibitem{2020Zhang}
J.~Zhang, X.~Yu, and K.~B. Letaief, ``Hybrid beamforming for {5G} and beyond
  millimeter-wave systems: A holistic view,'' \emph{IEEE Open J. Commun. Soc.},
  vol.~1, pp. 77--91, Jan. 2020.

\bibitem{2022Yu}
W.~Yu, Y.~Shen, H.~He, X.~Yu, J.~Zhang, and K.~B. Letaief, ``Hybrid far- and
  near-field channel estimation for {THz} ultra-massive {MIMO} via fixed point
  networks,'' in \emph{Proc. IEEE Global Commun. Conf. (GLOBECOM)}, Rio de
  Janeiro, Brazil, Dec. 2022.

\bibitem{2017Borgerding}
M.~Borgerding, P.~Schniter, and S.~Rangan, ``{AMP}-inspired deep networks for
  sparse linear inverse problems,'' \emph{IEEE Trans. Signal Process.},
  vol.~65, no.~16, pp. 4293--4308, Aug. 2017.

\bibitem{2022He}
H.~He, R.~Wang, W.~Jin, S.~Jin, C.-K. Wen, and G.~Y. Li, ``Beamspace channel
  estimation for wideband millimeter-wave {MIMO}: A model-driven unsupervised
  learning approach,'' \emph{IEEE Trans. Wireless Commun.}, to appear.

\bibitem{2022ShenTWC}
Y.~Shen, J.~Zhang, S.~Song, and K.~B. Letaief, ``Graph neural networks for
  wireless communications: From theory to practice,'' \emph{IEEE Trans.
  Wireless Commun.}, to appear.

\bibitem{2021Zheng}
X.~Zheng and V.~K.~N. Lau, ``Online deep neural networks for {mmWave} massive
  {MIMO} channel estimation with arbitrary array geometry,'' \emph{IEEE Trans.
  Signal Process.}, vol.~69, pp. 2010--2025, Mar. 2021.

\bibitem{2022Zou}
Q.~Zou and H.~Yang, ``A concise tutorial on approximate message passing,''
  \emph{arXiv preprint arXiv:2201.07487}, 2022.

\bibitem{1981Stein}
C.~M. Stein, ``Estimation of the mean of a multivariate normal distribution,''
  \emph{Ann. Statist.}, pp. 1135--1151, Nov. 1981.

\bibitem{2021Gallyas-Sanhueza}
A.~Gallyas-Sanhueza and C.~Studer, ``Blind {SNR} estimation and nonparametric
  channel denoising in multi-antenna {mmWave} systems,'' in \emph{Proc. IEEE
  Int. Conf. Commun. (ICC)}, Montreal, Canada, Jun. 2021.

\bibitem{2022Cao}
R.~Cao, H.~He, X.~Yu, J.~Zhang, S.~Song, Y.~Gong, and K.~B. Letaief, ``Belief
  propagation for near-field cooperative localization and tracking in {6G}
  vehicular networks,'' in \emph{Proc. IEEE Int. Mediterr. Conf. Commun. Netw.
  (MeditCom)}, Athens, Greece, Sept. 2022.

\bibitem{2017Ma}
J.~Ma and L.~Ping, ``Orthogonal {AMP},'' \emph{IEEE Access}, vol.~5, pp.
  2020--2033, Jan. 2017.

\bibitem{2016Metzler}
C.~A. Metzler, A.~Maleki, and R.~G. Baraniuk, ``From denoising to compressed
  sensing,'' \emph{IEEE Trans. Inf. Theory}, vol.~62, no.~9, pp. 5117--5144,
  Sept. 2016.

\bibitem{yu2022adaptive}
W.~Yu, Y.~Shen, H.~He, X.~Yu, S.~Song, J.~Zhang, and K.~B. Letaief, ``An
  adaptive and robust deep learning framework for {THz} ultra-massive {MIMO}
  channel estimation,'' \emph{arXiv preprint arXiv:2211.15939}, 2022.

\bibitem{2007Blu}
T.~Blu and F.~Luisier, ``The {SURE-LET} approach to image denoising,''
  \emph{IEEE Trans. Image Process.}, vol.~16, no.~11, pp. 2778--2786, Nov.
  2007.

\bibitem{2008Ramani}
S.~Ramani, T.~Blu, and M.~Unser, ``{Monte-Carlo Sure}: A black-box optimization
  of regularization parameters for general denoising algorithms,'' \emph{IEEE
  Trans. Image Process.}, vol.~17, no.~9, pp. 1540--1554, Sept. 2008.

\bibitem{2015Chen}
G.~Chen, F.~Zhu, and P.-A. Heng, ``An efficient statistical method for image
  noise level estimation,'' in \emph{Proc. IEEE Conf. Comput. Vis. Pattern
  Recog. (CVPR)}, Boston, MA, USA, Jun. 2015.

\bibitem{2016He}
K.~He, X.~Zhang, S.~Ren, and J.~Sun, ``Deep residual learning for image
  recognition,'' in \emph{Proc. IEEE Conf. Comput. Vis. Pattern Recog. (CVPR)},
  Las Vegas, NV, USA, Jun. 2016.

\bibitem{2016Ba}
J.~L. Ba, J.~R. Kiros, and G.~E. Hinton, ``Layer normalization,'' in
  \emph{Proc. Adv. Neural Inf. Process. Syst. (NeurIPS)}, Barcelona, Spain,
  Dec. 2016.

\end{thebibliography}

\end{document}